\documentclass{PoS}

\title{G parity boundary conditions and $\Delta I = 1/2$, $K \rightarrow \pi \pi$ decays}

\ShortTitle{G parity boundary conditions and $\Delta I = 1/2$, $K \rightarrow \pi \pi$ decays}

\author{Changhoan Kim$^\dagger$\\
        Columbia University, USA\\
        E-mail: \email{ck364@phys.columbia.edu}}

\author{\speaker{Norman Christ}
        \thanks{This work was partially supported by DOE grant 
        \#DE-FG02-92ER40699 and by the RIKEN BNL Research Center.}\\
        Columbia University, USA\\
        E-mail: \email{nhc@phys.columbia.edu}}

\author{{RBC and UKQCD collaborations}}

\abstract{The use of G-parity boundary conditions
to compute $\Delta I = 1/2$, $K \rightarrow \pi \pi$ 
decays is reviewed and a method to consistently treat
both the pions and kaon in full QCD proposed.  This
approach creates a physical, final-state, pion momentum 
using a 3 fm box and avoids statistical noise coming 
from pions with smaller momentum.}

\FullConference{The XXVII International Symposium on Lattice Field Theory\\
		 July 26-31, 2009\\
		 Peking University, Beijing, China}

\def\cancel#1#2{\ooalign{$\hfil#1\mkern1mu/\hfil$\crcr$#1#2$}}
\def\slash#1{\mathpalette\cancel{#1}}

\begin{document}

The violation of CP symmetry in the two pion decays of the $K$ meson 
offers an important opportunity to uncover new sources of CP violation 
beyond those predicted by the standard model.  Of special interest is 
the direct CP violation parameterized by $\epsilon^\prime$ which is now 
experimentally determined on the 10\% level and sensitive to possible 
new phenomena on the TeV scale.  An accurate calculation of this quantity 
within the standard model requires the evaluation of matrix elements of 
four-Fermi operators between $K$ and $\pi-\pi$ states.  Such matrix 
elements are within the reach of lattice QCD methods.  However, the 
presence of two particles in one of the states and the need to the 
evaluate ``vacuum'' or ``disconnected'' diagrams, in which the initial 
and final states are joined only by the exchange of gluons, make these 
calculations particularly difficult.

Here we focus on the difficulties of the $\pi-\pi$ final state, 
especially the $I=0$ state with vacuum quantum numbers.  An attractive 
approach to these two-pion states uses chiral perturbation theory to 
relate the two pion matrix elements of interest to simpler matrix 
elements between the $K$ meson and a single pion or the vacuum state.  
Unfortunately, recent results~\cite{Allton:2008pn,Li:2008kc} suggests 
that $SU(3)\times SU(3)$ chiral perturbation theory works poorly at the 
kaon energy leading to large systematic errors from this approach. 
Thus, calculation of actual $\pi-\pi$ matrix elements have become 
important.

\section{Overview of finite volume methods}

The methods of lattice QCD construct the eigenstates of the full
QCD Hamiltonian $H_\mathrm{QCD}$ by studying Green's function of 
interpolating operators at large separations in Euclidean time.  In 
this way the contribution of that eigenstate with the lowest eigenvalue 
of $H_\mathrm{QCD}$ is exponentially enhanced.  This results in the 
difficulty of Maiana and Testa~\cite{Maiani:1990ca} that when studying
$K \rightarrow \pi \pi$ decays the energy non-conserving matrix elements 
with the state of two pions at rest will be computed.

As is now well understood, this difficult can be diminished by exploiting
the finite volume in which lattice calculations are necessarily performed.
In finite volume the $\pi-\pi$ eigenstates of $H_\mathrm{QCD}$ are a series 
of discrete states with energies shifted in a known way from those of two 
free particles in a box by the $\pi -\pi$ interaction~\cite{Luscher:1986pf,
Luscher:1990ux}.  

However, working in finite volume also introduces a difficulty.  The finite 
volume eigenstates of $H_\mathrm{QCD}$ are necessarily mixtures of different 
angular momenta because the usual cubic box is asymmetric under rotations.  
Thus, the matrix element of a local weak operator $Q_i$ between the $K$ 
meson and finite-volume $\pi-\pi$ eigenstate, $\langle K| Q_i|\pi\pi\rangle$, 
is a product of the desired $l=0$ decay amplitude and the amplitude for 
finding this $l=0$ state within the normalized finite volume $\pi-\pi$ 
eigenstate which is a superposition of different values of $l$.  This 
problem has been solved by Lellouch and Luscher~\cite{Lellouch:2000pv}.  
The needed correction can be made to adequate accuracy once the $\pi-\pi$, 
$l=0$ scattering phase shift has been determined for the relevant isospin 
chanel and energies. 

Three methods that have been developed to deal with these finite volume 
states.  In the first, one tunes the linear size $L$ of an $L^3 \times T$ 
space-time volume so the $2\pi/L$ (or more accurately the quantized momentum 
which is determined by the $\pi-\pi$ scattering phase shift $\delta(I)_{l=0}$) 
so that the first excited $\pi-\pi$ state has the energy of the kaon.  
For a physical pion this requires $L=6$ fm. While appealingly simple, 
in this approach the state of interest, $|\pi\pi(p\approx 2\pi/L\rangle$, 
is in fact the third lowest energy state that will contribute to the 
Green's function being computed: both the vacuum (for $I=0$ and the 
$p\approx 0$ $\pi-\pi$ state will have lower energy.  Extracting the 
decay matrix element from such a three-exponential description of this 
Green's function time dependence will be difficult.

A second method, which results in the $\pi-\pi$ state possessing the 
lowest energy of all allowed states, gives the initial kaon a non-zero 
3-momentum.  The two final pions must also carry this momentum.  
The lowest energy $\pi-\pi$ state with $\vec p \ne 0$ will contain one 
pion which is nearly at rest and a second which carries the kaon 
momentum.  An energy conserving decay will result if the kaon and one 
final pion carry 740 MeV of momentum.  For a 3.5 fm box (too small to 
avoid large finite volume corrections at the physical pion mass) the 
kaon and pion momenta would need to be $\approx 2 \cdot 2\pi/L$.  Such 
large momentum amplitudes are expected to be noisy and difficult to 
compute.

This approach with a non-zero center (cm) of mass momentum has been explored
theoretically~\cite{Christ:2005gi,Kim:2005gf} and a first calculation 
shows encouraging results~\cite{Yamazaki:2008hg}.  Since the vacuum 
cannot carry momentum, the lowest energy state which contributes to 
such non-zero cm momentum correlators will be the $\pi-\pi$ state of 
interest, even for $I=0$.  This approach deserves further study.

A third approach, which is the topic of the remainder of this article, is the 
use of boundary conditions to eliminate the $p \approx 0$, $\pi-\pi$ state.
There are two techniques of interest.  The first, which can be applied to
only the $I=2$, $\pi-\pi$ state, imposes anti-periodic boundary conditions 
on one of the two flavors of light quarks making up the pions.  This use
of twisted boundary conditions with a twist angle $\theta=\pi$ was 
introduced in Ref.~\cite{Kim:2005gk} where they were called H-parity 
boundary conditions.  For the case of $\Delta I = 3/2$ decay, the $\pi-\pi$
final state will have $I=2$ and isospin symmetry can be used to
relate the matrix element of interest to a matrix element involving the
$|\pi^+\pi^+\rangle$ state.  With H-parity boundary conditions, the
$\pi^+$ meson will obey anti-periodic boundary conditions forcing the
pions in a $\pi^+-\pi^+$ state with zero cm mass momentum to have a
relative momentum of $\pm\pi/L$ in the direction orthogonal to the 2-d
face on which the boundary conditions are applied.  Since this $\pi^+-\pi^+$
state is the unique $\pi-\pi$ state with charge 2, the use of isospin 
breaking boundary conditions does not lead to unwanted mixing of the 
$I=0$ and $I=2$ final states.  Finally because all $\pi-\pi$ intermediate 
states must contain the valence quarks on which the boundary conditions 
have been imposed, such H-parity boundary conditions can be imposed on 
only the valence quarks, using a lattice ensemble generated with normal 
boundary conditions~\cite{Sachrajda:2004mi}.  This is an attractive 
method to compute the $\Delta I = 3/2$ kaon decay amplitude
~\cite{Kim:2004thesis,Lightman:2009xx}.

A second type of boundary condition which can be imposed to insure
that the final-state pions carry non-zero momentum is G-parity boundary 
conditions~\cite{Kim:2002np}.  Since all three pions are odd under 
G-parity, these boundary conditions imply that each pion must have 
$p_i \approx \pm\pi/L$ when these boundary conditions are imposed on 
the face perpendicular to the $i^{th}$ direction.  We now discuss these 
boundary conditions in more detail.
 
\section{G-parity boundary conditions for the pions}

Recall that the G-parity transformation composes the charge conjugation 
operator with an isospin rotation about the $y$-direction.  This 
transformation commutes with the three generators of isospin and changes 
the sign of an iso-triplet whose third component, in our case the 
$\pi^0$, is charge conjugation even.  On the level of the quark fields 
creating $u$ and $d$ quarks, $(\overline{u},\overline{d})$ G-parity has 
the following action:
\begin{equation}
G\left(\begin{array}{c} \overline{u} \\ \overline{d} \end{array}\right)G^{-1}
  = 
\left(\begin{array}{c} C d^{\rm T} \\ -C u^{\rm T} \end{array}\right)
\label{eq:G_def}
\end{equation}
where $C$ is the $4\times 4$ charge conjugation matrix which obeys 
$C \gamma^\mu C^{-1} = -\left(\gamma^\mu\right)^{\rm T}$.  Here and in 
Eq.~\ref{eq:G_def} above, the superscript $T$ indicates the transpose 
of a $4 \times 4$ and a $1 \times 4$ matrix respectively.  Such 
G-parity boundary conditions are ideal for our problem, respecting 
isospin symmetry while yielding a lowest energy state of the two 
pions of the form:
\begin{equation}
 E_{\pi\pi} \approx \sqrt{n\left(\frac{\pi}{L}\right)^2 + m_\pi^2}.
\end{equation}
where $n =0,$ 1, 2, 3 is the number of spatial boundaries on which the
G-parity condition is imposed.  The appearance of $\pi/L$
rather than $2\pi/L$ means that smaller energies are
accessible.

Two difficulties must be overcome when implementing these boundary 
conditions.  First the gauge links that cross a boundary across 
which G-parity is imposed connect quark fields that transform as 
SU(3) color triplets and anti-triplets.  Thus, these links must 
transform under a gauge transformation V(x) as
$U_\mu(x) \rightarrow V(x)U_\mu(x)V(x+(1-L)\hat e_\mu)^{\rm T}$ assuming 
that the site $x$ is adjacent to a boundary which the link $U_\mu(x)$
crosses and that $L$ is the lattice size in the $\mu$ direction.  
This modified transformation law requires a modified gauge 
action for all plaquettes which straddle this boundary.  For $x$ and
$\mu$ as above and $\nu \ne \mu$ we must use 
tr$\left(U_\mu(x) U_\nu(x+(1-L)\hat e_\mu)^* U_\mu(x+\hat e_\nu)^{-1}
U_\nu(x)^{-1}\right)$.  Of course, the required change in the gauge
action can be made and the resulting theory will still be 
translationally invariant provided the translation operation is
generalized to include replacing some gauge links with their complex
conjugate.  This altered gauge action requires that a new ensemble
of gauge field be generated for each assignment of G-parity boundary
conditions.

The second complexity is the presence of charge conjugation in the 
definition of G-parity.  Typically a Dirac operator that includes 
such a charge conjugation will be represented by a path integral 
which evaluates to Pfaffian~\cite{Zinn-Justin:1989} rather than a 
more familiar determinant.  However, for our two-flavor case there 
is no direct coupling between a Grassmann integration variable and 
itself. For example, $u$ couples to $\overline{d}$ which then couples 
to $-u$ as one traverses the lattice twice in a direction 
perpendicular to a face across which G-parity conditions are imposed.  
Thus, we can view the $u$ quark as a Grassmann variable defined on 
a doubled volume where the $u$ degree of freedom on the extention
of the original volume actually equals $\overline{d}$.  The result
is a standard theory of a single flavor of quark obeying anti-periodic
boundary conditions on a doubled lattice volume.  While this prevents
the second quark from being represented by a simple change of an
$N_f$ factor appearing in the evolution algorithm from 1 to 2, the 
only real cost is a factor of two in the lattice volume that must be 
studied.  This added factor of two cost remains as G-parity is 
imposed in additional directions.  However, the resulting geometry
becomes more complex than an $L \times L \times 2L$ three-volume 
with simple boundary conditions relating the values of the Grassmann 
$u$ variable on opposite faces.

Thus, G-parity boundary conditions can be imposed on the $u$ and
$d$ quarks with no difficulty beyond the doubled space-time 
volume.  Initial numerical simulations have uncovered no serious
problems beyond enhanced finite-volume effects that come from 
the possible binding of an isolated quark to its own, distant, 
charge-conjugate image.    

\section{G-parity boundary conditions for the kaon}

However, we must now decide how to treat the strange quark.  The
imposition of charge conjugation boundary conditions on the gauge
field implies that the strange quark cannot obey standard periodic
or anti-periodic boundary conditions.  There are two natural 
options to consider.  The strange quark could obey charge-conjugate
boundary conditions.  Alternatively the strange quark could be made 
part of an artificial pair of degenerate quarks which transform as 
an isospin doublet and obey G-parity boundary conditions.  We 
consider each of these possibilities in turn.

Let us first impose C boundary conditions on a single species of 
strange quark.  Begin with the standard Grassmann action for the
strange quark and then rewrite one half of that action by reversing
the order of the $s$ and $\overline{s}$ fields:
\begin{equation}
\overline{s} \slash{D} s = 
\frac{1}{2} \left(\begin{array}{cc} s^{\rm T}          & \overline{s}  \end{array}\right)
      \cdot \left(\begin{array}{cc} 0            & -\slash{D}^{\rm T}  \\
                                    \slash{D}    & 0             \end{array}\right)
      \cdot \left(\begin{array}{c}  s                            \\
                                    \overline{s}^{\rm T}                \end{array}\right)
=\frac{1}{2}
  \Psi^{\rm T} \cdot \left(\begin{array}{cc} 0         & -\slash{D}^{\rm T}  \\
                                    \slash{D}    & 0             \end{array}\right)
         \cdot \Psi 
\mbox{~~~where~~~} \Psi =  \left(\begin{array}{c} s                            \\
                                    \overline{s}^{\rm T}                \end{array}\right).
\label{eq:C-BCs_I}
\end{equation}
The standard boundary conditions appear as off-diagonal terms in the Dirac operator 
$\slash{D}$.  We can change these boundary conditions to charge conjugation boundary 
condition by removing this off-diagonal term from $\slash{D}$ and putting it into 
the diagonal blocks labeled as zero in Eq.~\ref{eq:C-BCs_I}.  If the Dirac operator 
without this diagonal term is written as $\slash{D}\,^\prime$ this change in boundary
conditions will result in a new action which can be written schematically as:

\begin{equation}
 \Psi^{\rm T} \cdot \left(\begin{array}{cc} 
              \left[\begin{array}{cc} 0 & \Delta \\
                              -\Delta^{\rm T} & 0       \end{array} \right]
                                                & -{\slash{D}\,^\prime}^{\rm T} \\
  \slash{D}\,^\prime & \left[\begin{array}{cc} 0 & \Delta^\prime \\
                           -{\Delta^\prime}^{\rm T} & 0       \end{array} \right] 
                  \end{array}\right) \cdot \Psi 
\quad\rightarrow\quad
 \overline{\Theta} \cdot \left(\begin{array}{cc} 
              \left[\begin{array}{cc} 0 & \Delta \\
                              -\Delta^{\rm T} & 0       \end{array} \right]
                                                & -{\slash{D}\,^\prime}^{\rm T} \\
  \slash{D}\,^\prime & \left[\begin{array}{cc} 0 & \Delta^\prime \\
                            -{\Delta^\prime}^{\rm T} & 0       \end{array} \right] 
                  \end{array}\right) \cdot \Theta .
\label{eq:C-BCs_II}
\end{equation}
where the off-diagonal term $\Delta$ implements the coupling of $s(x,y,z=L-1,t)$ 
and $s^{\rm T}(x,y,z=0,t)$ for the case of charge conjugate boundary conditions 
in the $z$-direction and a lattice with $L$ sites in that direction.  Similarly  
$\Delta^\prime$ connects $\overline{s}^{\rm T}(x,y,z=L-1,t)$ and 
$\overline{s}(x,y,z=0,t)$.

The operator on the left side of the expression~\ref{eq:C-BCs_II} is the 
desired Dirac operator with charge conjugation boundary conditions.  Grassmann
integrals with this action and the integrand $\Psi(x) \Psi(y)^{\rm T}$ will give
the inverse of the Dirac operator obeying these boundary conditions times
the Pfaffian of that operator.  A practical way to evaluate that Pfaffian is
indicated by the term on the right of Eq.~\ref{eq:C-BCs_II} where the number
of independent fermion fields has been doubled so that the single field $\Psi$
has been replaced by two fields $\Theta$ and $\overline{\Theta}$.  This second
action is entirely standard and would yield a normal determinant, the square
of the Pfaffian of interest.  The resulting Dirac operator is defined in a 
volume doubled in the $z$-direction and obeying periodic boundary conditions 
in that doubled direction.  Using the usual rational hybrid Monte Carlo method, 
we could easily perform a dynamical simulation using the square root of this 
usual determinant and recover the desired Pfaffian.  If domain wall fermions 
are used with non-zero fermion mass, this determinant is guaranteed to be 
positive so its square root is well defined.  Given the connectivity of the 
gauge field integration volume, we can then choose a consistent sign for this 
square root which will be valid throughout that integration volume.

While it is encouraging that charge conjugation boundary conditions are
practical to implement, they will not solve our problem.  For example, a $K^0$ 
meson, created from these strange and light quarks, which is an eigenstate
under translation by $L$ in the $z$-direction will have the from 
$(\overline{s} d \pm i s \overline{u})/\sqrt{2}$ where, from the perspective
of the doubled lattice volume in the $z$-direction, the left-hand term 
describes the particle for $0 \le z < L$ and the right-hand term applies when
$L \le z < 2L$.  Unfortunately, under this translation by $L$, these states 
have eigenvalue $\pm i$ and therefore carry momentum $\pm \pi/2L$ so that a 
momentum conserving 2 pion decay is not possible.  This should be expected 
since $s(x)$ is even when translated in the $z$-direction by $2L$ while the 
light quark field is odd, implying that the K meson will satisfy anti-periodic 
boundary conditions in this expanded $2L$ volume and carry momentum 
$\pm \pi/2L$.  

Note that the odd mixture of particles making up the K meson does not create 
a problem.  By using a weak operator with the correct particle content, 
we can insure that only the physical $\overline{s} d$ part of the initial 
state will contribute.  The effect of the unusual mixture in this initial 
state is only on its normalization, introducing a factor of $1/\sqrt{2}$ 
which can be easily be removed.  In the doubled-volume language, such a 
physically correct choice for the weak operator must involve fields in
only one half of the doubled volume which are therefore not translationally
invariant so that conservation of momentum must be imposed by the (here
impossible) choice of initial and final states.

The second alternative of introducing a fictitious iso-doublet which is made
up of the strange quark and a second ``charm'' quark, $(c, s)$ avoids
the non-zero momentum problem discovered above.  Now both quarks in
a generalized initial kaon state will change sign under translation 
through $2L$ so a state with zero momentum becomes possible.  The 
ground state which contains the desired $\overline{s}d$ component will be
\begin{equation}
|K^0\rangle = \frac{1}{\sqrt{2}}\left(\overline{s}d + c\overline{u}\right). 
\end{equation}
This state is translationally invariant and consistent with the boundary
conditions permitting it to carry zero momentum as required.

With this choice of boundary conditions for the strange quark we have
effectively doubled the number of flavors in the strange quark sector.
Directly using the determinant of this doubled Dirac operator as the
weight in the QCD path integral would be incorrect because QCD has one 
not two flavors of strange quark.  Apparently the best solution to this
problem is to use the square root of this determinant.  This square root
will correspond to the correct number of flavors but will add non-locality
since such a square root cannot be realized by a local Grassmann path
integral.  Of course, in contrast with the rooting used with staggered
fermions, any effects of this non-locality will disappear in the limit 
of large volume.  The Dirac operator in question differs from the 
doubled Dirac operator obeying charge conjugation boundary conditions 
on the right side of Eq.~\ref{eq:C-BCs_II} by boundary terms and the 
determinant of this doubled operator obeying charge conjugation boundary
conditions is the square of a positive Pfaffian which is appropriately
local.

One might also worry that taking such a square root introduces a 
miss-match between the treatment of the valence and sea quarks.  
Again such non-unitary effects are expected to be exponentially 
suppressed for the case of single-particle states such as our 
K meson in which there is no mixing between the valence and sea 
quarks~\cite{Sachrajda:2004mi}.\footnote{It is interesting to note 
that this situation is very similar to the use of a superposition 
of propagators obeying periodic and anti-periodic boundary 
conditions in the time direction, often done to reduce finite volume 
effects.  The eigenvectors of the Dirac operator defined on the 
doubled lattice divide into states either periodic and anti-periodic
under translation through the original time extent $T$.   Thus, the
determinant of the Dirac operator defined on the doubled lattice
is the product of the determinants of the Dirac operators defined
on the original lattice obeying periodic and anti-periodic boundary 
conditions.  Using the correct number of flavors would require taking 
the square root of this determinant which is then also not quite a 
perfect square.} 

\section{Conclusions}

The use of G-parity boundary conditions for the up and down quarks and 
for an iso-doublet made of degenerate strange and fictitious 
charm quarks allows an accurate description of $K \rightarrow \pi\pi$
decays in which the finte-volume energy of the two-pion final state 
is quantized and the unphysical state with approximately zero relative 
momentum forbidden.  This approach is computationally demanding, 
requiring a new set of gauge configurations for each choice of boundary 
conditions.  However, for a computationally difficult problem such as 
posed by the $\pi-\pi$ state with $I=0$, the generation of the gauge
configurations may not be the dominant cost and this approach may 
yield better-controlled errors than the competing method of using a 
740 MeV center of mass momentum.  

\providecommand{\href}[2]{#2}\begingroup\raggedright\endgroup

\end{document}